# Superconducting Topological Surface States in Non-centrosymmetric Bulk Superconductor PbTaSe$_2$

Syu-You Guan[1,2]§, Peng-Jen Chen[1,2,3]§, Ming-Wen Chu[4], Raman Sankar[2,4], Fangcheng Chou[4], Horng-Tay Jeng[2,5]*, Chia-Seng Chang[1,2]*, and Tien-Ming Chuang[2]*

[1] Department of Physics, National Taiwan University, Taipei 10617, Taiwan
[2] Institute of Physics, Academia Sinica, Nankang, Taipei 11529, Taiwan
[3] Nano Science and Technology Program, Taiwan International Graduate Program, Academia Sinica, Taipei 11529, Taiwan and National Taiwan University, Taipei 10617, Taiwan
[4] Center for Condensed Matter Sciences, National Taiwan University, Taipei 10617, Taiwan
[5] Department of Physics, National Tsing Hua University, Hsinchu 30013, Taiwan
§ These authors contributed equally to this work.
*e-mail: jeng@phys.nthu.edu.tw; jasonc@phys.sinica.edu.tw; chuangtm@phys.sinica.edu.tw

**Abstract**

1    The search for topological superconductors (TSCs) is one of the most urgent contemporary problems in condensed matter systems. TSCs are characterized by a full superconducting gap in the bulk and topologically protected gapless surface (or edge) states. Within each vortex core of TSCs, there exist the zero energy Majorana bound states, which are predicted to exhibit non-Abelian statistics and to form the basis of the fault-tolerant quantum computation. So far, no stoichiometric bulk material exhibits the required topological surface states (TSSs) at Fermi level ($E_F$) combined with fully gapped bulk superconductivity. Here, we report atomic scale visualization of the TSSs of the non-centrosymmetric fully-gapped superconductor, PbTaSe$_2$. Using quasiparticle scattering interference (QPI) imaging, we find two TSSs with a Dirac point at E≅1.0eV, of which the inner TSS and partial outer TSS cross $E_F$, on the Pb-terminated surface of this fully gapped superconductor. This discovery reveals PbTaSe$_2$ as a promising candidate as a TSC.

**Introduction**

2    The helical spin-polarized electrons on the surface of topological insulators (TIs) can host exotic excitations such as Majorana fermions (*1-6*) when TSS goes superconducting. Two primary proposals for realizing such a TSC are (i) spin-triplet pairing bulk superconductivity (*1, 2, 4*) and (ii) superconducting topological surface states (TSSs) induced by proximity effect through a s-wave superconductor (*3, 7*). A chiral p-wave superconductor is a natural choice as a bulk TSC but electrons rarely form spin-triplet Cooper pairs in real materials. The order parameter of the prime candidate Sr$_2$RuO$_4$ remains under debate (*8, 9*) while the inhomogeneity within superconducting doped TIs hinders the confirmation of Majorana bound states (MBS) at the vortex cores (*10*). Therefore, searching for a TSC remains one of the grand challenges in current TI research. Non-centrosymmetric superconductors (NCSs) offer one of the most promising routes to realize TSCs (*11, 12*). Anti-symmetric spin orbit coupling (SOC) due to the broken inversion symmetry can lift spin degeneracy, giving rise to topological phases and also allowing the mixing between spin-triplet and spin-singlet pairing channel. Hence, a NCS with dominating spin-triplet pairing can become a bulk TSC, whereas for a NCS with dominating spin-singlet pairing in the bulk, it could still lead to a TSC on the surface if the fully gapped superconductivity were induced on TSS. A NCS with superconducting TSS is greatly desired to provide the required platform for TSC.

**3**     Previously, non-centrosymmetric BiPd is shown to exhibits a conventional BCS s-wave superconducting state and TSS above $T_C$; however, TSSs in BiPd are absent at $E_F$, unable to become a TSC (*13, 14*). Similarly, the observation of TSSs in superconducting centrosymmetric β-PdBi$_2$ also attracts great interest (*15*). Parity mixing of pair potential may occur near the surface since the inversion symmetry is broken at the surface. Furthermore, the presence of TSSs at $E_F$ in the normal state of β-PdBi$_2$ by ARPES (*15*) and QPI imaging (*16*), combined with the observation of fully gapped superconductivity in the bulk (*16-18*), makes β-PdBi$_2$ a great candidate for TSC. The zero energy bound states are absent at vortex cores of single crystal samples (*17*); however, they have been observed on the epitaxial β-PdBi$_2$ thin films (*19*), indicating the possible existence of MBS. Such a difference could be due to defects scattering or different surface/sample preparation, calling for further experiments. Recently, PbTaSe$_2$ has emerged as a strong TSC candidate (*20*). ARPES measurements observe topological nodal lines around $\bar{K}$ and two surface bands (*21*). However, the existence of a Dirac point and the spin texture of these two surface bands has not yet been confirmed because the Dirac point energy ($E_D$) is located at several hundred meV above $E_F$; hence, not accessible to ARPES. For the bulk superconducting state, although all recent bulk measurements indicate that PbTaSe$_2$ is likely a conventional BCS s-wave superconductor (*20, 22-25*), it has been proposed that the superconductivity on TSSs induced by proximity effect in PbTaSe$_2$ can be stronger than that in TI/s-wave superconductor heterostructures (*26*). Therefore, it is of critical importance to confirm the existence of a superconducting TSS and to investigate the potential signature of any MBS in PbTaSe$_2$.

4      In this study, our density functional theory (DFT) calculations reveal two TSSs with opposite helical spin polarization exist within the bulk gap at $\bar{\Gamma}$, touching at $E_D \cong E_F+1.0eV$. The inner TSS is fully isolated from the nearby bulk states at $E_F$ while the outer TSS heavily merges with bulk states at $E_F$. The characteristic wavevectors and their energy dispersion measured by our quasiparticle scattering interference (QPI) imaging at the normal state show remarkable agreement with our theoretically calculated spin-polarized TSS band structures on PbTaSe$_2$. The tunneling spectrum at T=0.26K reveals a full superconducting gap, indicating TSS becomes superconducting with bulk. Our vortex imaging also observes the zero energy bound states at superconducting vortex cores. Our results demonstrate stoichiometric PbTaSe$_2$ is a strong TSC candidate because its TSS can host Cooper pairs and becomes superconducting, providing a quintessential playground for nontrivial band topology and superconductivity.

**Results**

5      Our PbTaSe$_2$ single crystals were grown by chemical vapor transport technique and exhibit a bulk superconducting transition at $T_C$~3.8K (details are described elsewhere (*23*)). The crystal structure of PbTaSe$_2$ consists of alternating stacking of hexagonal 1H-TaSe$_2$ and Pb layers (Fig.1A). Within TaSe$_2$ layer, Ta atoms are suggested to reside at the same position within unit cell (space group: $P\bar{6}m2$) (*20*), breaking the inversion symmetry. This is confirmed by our atomically-resolved high-angle annular dark-field (HAADF) imaging (Fig.1B and 1C) in an aberration-corrected STEM, which unambiguously rules out the other possible crystalline structures reported in PbTaSe$_2$ (*23, 27*). Based on our experimentally confirmed crystal structure, we carry out detailed DFT calculations for the bulk band structure and topological invariant. Our results on the bulk band structures with and without SOC (Fig.1D and 1E, respectively) demonstrate that SOC plays an important role in PbTaSe$_2$, as previously reported (*20, 26*). A continuous gap throughout the Brillouin zone is present due to the gap opening by SOC, which is crucial for the topological property of PbTaSe$_2$. Two groups of bands that lie below and above this continuous gap are completely separated from each other. In this case, the Z$_2$ topological invariant is well-defined regardless of the metallicity of the system (*28*). We use the method proposed by R. Yu *et al.* (*29*) to compute the Z$_2$ invariant of the states below the continuous gap. The calculated Z$_2$=1 reveals the necessity of a TSS of PbTaSe$_2$. The TSS is further evidenced by the calculated surface band structure (Fig.1F). The orbital decomposition of the surface Pb and Ta atoms reveals that two TSSs stem from Pb-*p* orbitals while Ta-*d* orbitals also contribute significantly around the zone center. Apparently, the topological surface bands do not exhibit linear dispersion even in the vicinity of the crossing point at $E_D \sim E_F+1.0$eV at $\bar{\Gamma}$. Additional quadratic or cubic terms are required to describe the behaviors of Dirac electrons. Hereafter, we use "Dirac point" for brevity despite the non-linearity. We note that the outer TSS which represents the upper part of Dirac cone is present at $E_F$ due to this non-linearity and it strongly merges with the bulk states below $E \sim 300$meV along $\overline{\Gamma K}$ and below $E \sim -750$meV along $\overline{\Gamma M}$, evident from the much smaller contribution of surface Pb atoms (gray areas in Fig.1F). Consequently, only the inner TSS is well-separated from the bulk states at $E_F$, consistent with Fermi surface revealed by ARPES (*21*). This is important because the odd number of TSS at Fermi surface is required for the stability of MBS in TSC-related theoretical proposals (*1, 2, 6, 7*). The helical spin states of TSSs can be seen from the spin-decomposed band structures (Fig.S2). From our DFT results, the TSS of PbTaSe$_2$ is confirmed by the $Z_2$ analysis and the fingerprints revealed in the surface band structure, consistent with a previous report (*26*). Experimentally, ARPES has been very successful to resolve the band structures in TIs. However, in the case of PbTaSe$_2$, these calculated TSSs and Dirac point are located above $E_F$ at $\bar{\Gamma}$, which is not accessible to ARPES. Thus, momentum sensitive techniques that can probe both occupied and unoccupied states are required to resolve this issue.

**6** SI-STM is an ideal technique to investigate the calculated TSSs and superconductivity in PbTaSe$_2$ by local density of state (LDOS) mapping in real space and QPI imaging in momentum space. Previous SI-STM and theoretical studies on topological materials have demonstrated that the band structures of TSS and the signature of its spin order can be visualized by QPI imaging (*30-37*). Our samples are cleaved in ultrahigh vacuum at T<20K before SI-STM measurements. The cleavage plane is between Pb and TaSe$_2$ layers. Topographic image (Fig.2A) shows a typical cleaved surface with several atomically flat terraces up to few hundred nm. The step height between terraces is ~0.3nm (Fig.2B), comparable with the shortest distance between Pb and Se. Atomic resolution STM topographic images taken at different terraces show two distinct surfaces. The lower area in Fig. 2A shows a hexagonal lattice (a~0.34nm) with atoms of 2×2 superstructure residing in between hexagons, consistent with TaSe$_2$ layer (Fig.2C). We thus deduce this to be Se-terminated TaSe$_2$ layer. Our calculation shows that the 2×2 superstructure arises from the tiny shift in Ta atomic positions (Fig.S3). In contrast, topographic image taken at the higher area shows a hexagonal lattice (a~0.34nm) with strong LDOS modulation from electronic surface states (Fig. 2D and 2E). Because Pb-*p* orbitals dominate TSSs near the bias voltage of topographic images, we identify this as Pb-terminated surface. We further investigate LDOS in these two types of surfaces by differential tunneling conductance (dI/dV) measurements. Typical normalized spectra taken at Se-terminated and Pb-terminated surface at T=0.26K are shown in Fig. 2G and 2H, respectively. The spectrum acquired from a Se-terminated surface shows its metallic nature with increasing DOS below $E_F$, which we associate with the contribution from bulk electronic states. In comparison, the spectrum taken at a Pb-terminated surface exhibits almost identical DOS but with two pronounced peaks at E~0.5eV and E~0.95eV, which are the signature of a van Hove singularity due to the saddle point of TSSs near E~0.5eV and the local maxima at $E_D$ (Fig.1F), respectively, indicating the existence of TSSs. Our tunneling spectra are also in excellent agreement with our calculated surface-projected PDOS (Fig.2G, 2H), affirming our surface identification. Pb-*p* orbitals dominate these TSSs over the energy range of $E_F\pm0.5$eV (Fig.1F). Even for the states close to the Dirac point where Ta-d orbitals become significant, Pb-*p* orbitals still have considerable contributions. Therefore, our QPI imaging will focus on Pb-terminated surface to determine the TSSs band structure.

**7** To resolve the ***k***-space electronic structures by QPI imaging, we first measure energy resolved differential conductance maps dI/dV(***r***, E=eV) on a Pb-terminated surface in the normal state (T=6.0K) with energy range from E=-0.20eV to 1.05eV by using our home-built cryogenic UHV SI-STM (Methods). In Fig.3A-3F, we show simultaneous normalized dI/dV(***r***, E) maps of six energies. It is apparent that the electron standing wave exhibits near-circular interference patterns at E≳300meV (Fig.3A-3D), whereas six fold symmetric patterns are visible at E≲300meV (Fig.3E-3F). Then we analyze Fourier transform of dI/dV(***r***, E) maps as function of energy and we show dI/dV(***q***, E) maps at corresponding energies (Fig.3G-3L). A ***q***-vector (***q***$_1$) with sharp circular closed contour is observed at E≳500meV (Fig.3G-I). Its length decreases with increasing energy and vanishes above E~1.0eV. From the helical spin texture of two TSSs and their energy dispersion, we deduce that ***q***$_1$ is the scattering vector connecting the opposite sides of two TSSs, which have the same in-plane spin orientation (Fig.4A and Fig.S4). When energy decreases from E~500meV to E~300meV, ***q***$_1$ becomes blurry and elongated (Fig.3J), which coincides when the dispersion of both TSSs becoming flat and also the out-of-plane spin components becoming larger, lifting the in-plane backscattering protection. When the energy is below E~300meV, another $q$-vector ($q_2$) is observed. Furthermore, both $q_1$ and $q_2$ disperse only along the $\overline{\Gamma M}$ direction with six-fold symmetry below E~300meV (Fig.3K-L) because warping effect on both TSSs (but helical spin polarization is still preserved) opens new scattering channels and the outer TSS overlapping with bulk states near $\overline{K}$ suppresses QPI along the $\overline{\Gamma K}$ direction (Fig.4B and Fig.S4). Relatively isotropic $q$-vector ($q_3$) is also observed near $E_F$ (Fig.3K-3L), which is due to intraband scattering from a bulk band at $\overline{\Gamma}$. In Fig. 3M-3R, simulated dI/dV($q$, E) images for these two TSSs at the corresponding energies appear in excellent agreement with our measurements. In Fig.4A and 4B, the scattering process of $q_1$ and $q_2$ are depicted on the constant energy contour plots of these TSSs with calculated spin orientation near $E_D$ and at $E_F$, respectively. The detail energy dispersion of three $q$-vectors extracted from our QPI data along $\overline{\Gamma M}$ depicted in Fig.4C and 4D appears in remarkable agreement with our band calculation and we can deduce $E_D$~1.0eV. Therefore, our QPI imaging demonstrates that the signature of two TSSs with opposite helical spin polarization and the Dirac point in the unoccupied states of PbTaSe$_2$, consistent with DFT calculations.

**8**     Next we study the superconducting state of PbTaSe$_2$ by temperature dependent and magnetic field dependent SI-STM measurements by our home-built $^3$He STM (Methods). The superconducting gap measured at T=0.26K does not show any spatial dependence on both Pb-terminated and Se-terminated surfaces (Fig.S5). The representative tunneling spectrum reveals the fully gapped superconductivity (Fig.5A). We then compare the tunneling spectrum taken by two STMs used in this study (Fig.S7). The only noticeable difference in LDOS above and below $T_C$ is the opening of a full superconducting gap at $E_F$, indicating the superconducting gap opens on TSSs together with the bulk counterpart. By fitting our data with an isotropic BCS s-wave gap, the temperature dependence of superconducting gap exhibits BCS-like behavior (Fig.5B, 5C). In addition, $2\Delta/k_BT_c$~3.4 from our data is close to the BCS ratio of 3.5 and different from that of 4.3 in bulk Pb (*38*) and monolayer Pb on Si(111) (*39*), consistent with weakly coupled s-wave superconductivity by bulk measurements (*20, 23, 24*). The observed superconducting gap at TSSs in the presence of helical spin polarization shows that PbTaSe$_2$ possesses the necessary ingredients for realizing TSC (*3, 7*).

9      A key question is then whether topological superconductivity indeed exists on the surface of PbTaSe$_2$. The existence of MBS at topological defects such as vortices would provide the definitive evidence by direct visualization of the vortex core states. Our vortex imaging at T=0.26K show a nearly perfect triangular Abrikosov lattice (Fig.5D) in a 550×550nm$^2$ field of view, from which we obtain the lattice constant of 239 nm with flux quanta $\Phi_0=2.0\times10^{-15}$Tm$^2$, consistent with theoretical lattice constant of Abrikosov vortices of 218 nm at 0.05T (Fig.S8). To understand the detailed electronic structures of a single vortex core, dI/dV($r$, E) maps were taken at several different magnetic fields. The vortex core exhibits a zero bias conductance peak (ZBCP), which can be the ordinary Caroli-de Gennes–Matricon (CdGM) bound state or MBS (Fig.5E). Although MBS is predicted to dominate at vortex cores (*26*), the thermal broadening at T=0.26K overlaps the small energy level spacing ($\Delta^2/E_F\sim10^{-7}$eV) of these vortex bound states, making it impossible to discriminate the CdGM bound state and MBS (*6*). Owing to the difficulty of identifying MBS by direct vortex imaging, the half-integer conductance quantization by point contact experiment, or thermal metal-insulator transition could provide alternative route to detect the signature of Majorana fermions if PbTaSe$_2$ were indeed a TSC (*4*). Recently it has been proposed that spatial dependence of the vortex core bound state can be used as the signature of MBS in Bi$_2$Se$_3$/NbSe$_2$ heterostructures, within which a spin polarized checkerboard pattern in LDOS is predicted (*40, 41*). The spatial dependence of LDOS along the line connecting two vortices is shown in Fig.5F. The ZBCP decays as the superconducting gap emerges with increasing distance from vortex core, which is identical with the spatial dependence of the vortex core state in topological trivial BiPd (*13*) and 2H-NbSe$_2$ (*42*). It will also be of great interest to investigate such phase-sensitive quasiparticle excitation spectrum in the vicinity of vortex cores in PbTaSe$_2$ by spin-polarized STM.

**Discussion**

10      Here we discuss the implication of our results on the realization of TSC in PbTaSe$_2$. Bulk measurements have demonstrated PbTaSe$_2$ to be a BCS s-wave superconductor with a full gap. The absence of a spin-triplet component in bulk superconductivity from the bulk measurements may be due to either the similar magnitude of the singlet and triplet component with the same sign (*12*) or the lack of strong electron correlations in PbTaSe$_2$ (*43*). On the scenario of the superconductivity on TSS induced by the proximity effect from bulk s-wave superconductivity, helical spin polarized electrons on the TSS can pair into $p_x+ip_y$ symmetry and bound Majorana fermions in the vortices (*3, 7*). Thus, it is also essential to understand the nature of superconductivity on the TSS. For TI/s-wave superconductor heterostructures, if the proximity-induced superconductivity reaches the TI surface, the induced gap is expected to be smaller than the bulk gap because the pairing amplitude decays quickly with increasing distance between the TI surface and the interface (*44*). In contrast, for an intrinsic superconducting TI with bulk s-wave pairing, the induced gap magnitude is shown to be similar with the bulk gap (*45*). If we fit our STS with a two full gaps model, we can obtain two gaps with comparable magnitudes (Fig. S6). A larger superconducting gap on the TSS may also be induced due to parity mixing of the pairing potential enhanced by surface Dirac fermions (*46*). Tunneling spectrum on superconducting $Cu_xBi_2Se_3$ and $Sr_xBi_2Se_3$ also do not show additional superconducting gap (*10, 47, 48*), which is proposed to be due to the doping evolution of Fermi surface (*46, 49*). Meanwhile, recent NMR (*50*) and specific heat (*51*) measurements on $Cu_xBi_2Se_3$ report rotational symmetry breaking in spin-rotation and superconducting gap amplitude, indicating a topological nontrivial superconductivity. Whether this also occurs in $PbTaSe_2$ requires further investigation. In principle, bound states at magnetic and non-magnetic impurities can be very helpful to determinate the pairing symmetry of unconventional superconductors (*52*). We emphasize that because no well-defined interface exists between TSS and bulk states in stoichiometric $PbTaSe_2$ and both states contribute to the measured tunneling current, which is evident from our QPI images, experimentally it is not straightforward to separate the contribution between bulk and TSS superconductivity in tunneling spectrum and realistic theoretical models are required for comparison. Further experiments such as band-selective superconducting gap mapping using QPI (*53*) or impurity-dependent spin-polarized QPI imaging (*54*) are more suitable to understand the superconductivity on TSS and bulk in $PbTaSe_2$.

**11**  In summary, our data unveil the nontrivial band topology of $PbTaSe_2$ in the normal state, showing two TSSs with opposite helical spin polarization touching at $E_D\sim 1.0eV$ at $\bar{\Gamma}$. The inner TSS is fully isolated from the nearby bulk states at $E_F$ while the outer TSS heavily merges with bulk states at $E_F$. The proximity induced fully gapped superconducting TSS is observed for the first time in a stoichiometric bulk material. This reveals that stoichiometric bulk $PbTaSe_2$ and closely related compounds hold promise to be TSC. $PbTaSe_2$ is especially so because it possesses all necessary ingredients for TSC as a consequence of the surface-bulk proximity effect. It is now imperative to understand the nature of the superconductivity of the TSS and to explore the novel topological superconducting phases towards future applications such as topological quantum computation.

## Materials and Methods

The DFT calculations are carried out using the Quantum Espresso (*55*) with norm-conserving pseudopotentials (*56*). Spin-orbit coupling (SOC) is included to study the topological properties. The k-meshes used to sample the BZ are 15×15×6 and 8×8×1 in the bulk and surface calculations, respectively. A slab consisting of 8 unit cell thickness (33 atomic layers) with both sides terminated at the layer of Pb is used in surface calculations. The energy cutoff of the plane wave expansion is 50 Ry. The lattice parameters from experiments are adopted in the calculations (a = 3.42 Å and c/a = 2.74).

The scanning transmission electron microscopy (STEM) investigations were conducted on a JEOL-2100F microscopy, operated at 200 keV and equipped with a CEOS spherical-aberration corrector, and the STEM high-angle annular dark-field (HAADF) imaging was acquired with the annular collection angle of 72~192 mrad. The electron-diffraction studies were performed on a FEI-Tecnai microscopy at 200 keV. All images were unprocessed and were taken at room temperature.

Scanning tunneling microscopy (STM) measurements: Single crystal $PbTaSe_2$ were cleaved at T<20K in ultrahigh vacuum and immediately inserted into the STM head for measurement. For QPI measurements in the normal state, we use a home-built cryogenic UHV SI-STM with base temperature of 1.6K, magnetic field up to 9T and hold time of ~10 days. For study of the superconducting state, we use a home-built $^3$He STM with base temperature of 0.25K, magnetic field up to 7T and hold time of ~24hours. Etched tungsten wires were used for STM tips. A standard ac lock-in technique was used for differential conductance measurements. QPI maps are unprocessed except only *q*=0 is suppressed to enhance the overall contrast of images.

## Supplementary Materials

Note S1. Crystal Structures and Electron Diffraction Patterns of $PbTaSe_2$.

Fig. S1. The electron-diffraction patterns along [001] and [110] projection.

Fig. S2. Spin-decomposed Surface Band Structures of $PbTaSe_2$.

Note S3. The 2×2 Superstructure on Se-terminated Surface.

Fig. S3. The 2×2 Superstructure on Se-terminated Surface.

Note S4. Verification of Helical Spin Polarization in Topological Surface States by QPI Imaging.

Fig. S4. The Simulation of QPI Images on Pb-terminated Surface.

**Acknowledgments**

We thank Ting-Kuo Lee, Sungkit Yip, J.C. Séamus Davis, Andreas W. Rost, Mohammad H. Hamidian, Hsin Lin, Wei-Feng Tsai, Tetsuo Hanaguri, Katsuya Iwaya, Dung-Hai Lee, Peter Wahl, Eun-Ah Kim, Mark H. Fischer, Ion Garate, Milan P. Allan, Wei-Cheng Lee, Wan-Ju Li, Sung-Po Chao and Tay-Rong Chang for helpful discussions and communications.

**General**: H.T.J. acknowledges National Center for High-Performance Computing (NCHC), Computer and Information Network Center (CINC) –NTU and National Center for Theoretical Sciences, Taiwan, for technical support.

**Funding:** This work is supported by Academia Sinica, National Taiwan University (NTU) and Ministry of Science and Technology. T.M.C is grateful for the support from Kenda Foundation.

**Author contributions:** S.Y.G and T.M.C. performed the SI-STM experiments and analyzed the data; R.S synthesized the samples; P.J.C and H.T.J. performed the theoretical calculations. M.W.C performed STEM experiments. F.C.C, H.T.J, C.S.C and T.M.C supervised the project. S.Y.G, P.J.C, H.T.J and T.M.C wrote the paper with input from other authors. The manuscript reflects the contributions of all authors.

**Competing interests:** The authors declare that they have no competing interests.
**Data and materials availability:** All data needed to evaluate the conclusions in the paper are present in the paper and/or the Supplementary Materials. Additional data available from authors upon request.

**Figure Captions**

**Fig. 1. Crystal structure and calculated band structure.** (**A**) Crystal structure of PbTaSe$_2$ and Brillouin zone of bulk (blue) and Pb-terminated surface (red). (**B**) The HAADF image projected along c-axis, overlapped with crystal structure signifying Ta (Pb/Se) atomic columns. (**C**) The HAADF image along [110] projection, revealing the characteristic Pb monolayer and 1H-TaSe$_2$ cages of PbTaSe$_2$. Calculated band structure of PbTaSe$_2$ (**D**) without SOC and (**E**) with SOC. The size of blue and red dots indicates the Pb-$p$ and Ta-$d$ orbital contributions, respectively. Large SOC splitting can be observed in Pb-$p$ and Ta-$d$ bands. The gray shaded area indicates the continuous gap. (**F**) The decomposed surface band structure on Pb-terminated surface of PbTaSe$_2$ shows the projection of Ta-$d$ and Pb-$p$ orbitals of the surface atoms. The gray circles indicated where the outer TSS merges into the bulk conduction bands.

**Fig. 2. Topographic Images and Tunneling Spectra.** (**A**) Topography of PbTaSe2, showing atomically flat terraces and step edges. (V=100mV and I=2pA). (**B**) The height profile deduced from the red line in (**A**). (**C**) The atomic resolution topographic image on Se-terminated surface (V=10mV and I=300pA). Inset shows Fourier transform of this topographic image. Yellow and red circle indicate the position of Bragg peaks and the peak of 2×2 superstructure, respectively. (**D**) Topography on Pb-terminated surface (V=10mV and I=100pA) shows strong LDOS modulation from QPI of topological surface states. (**E**) The atomic resolution topographic image on Pb-terminated surface (V=-10mV and I=300pA). (**F**) The FFT of topographic image in (**D**). Bragg peaks are marked by yellow circles, and the dispersive signals from QPI are observed (green ellipse). The normalized tunneling spectrum and calculated PDOS on (**G**) Se- and (**H**) Pb-terminated surface, respectively (T=0.26K, V=800mV, I=1nA and lock-in modulation=0.5mV). The superconducting gap near $E_F$ is not resolved clearly due to the large modulation in lock-in measurements.

**Fig. 3. Visualization of Quasiparticle Scattering Interference on Pb-terminated Surface.** (**A**)-(**F**) A sequence of normalized differential conductance dI/dV($r$, E)

maps taken at normal state (T=6K). The field of view (FOV) of each image is adjusted to access the region of interest in $q$-space. Larger FOV is required at higher energy near Dirac point to resolve the smaller $q$-vector. Scale bar in each image represents 10nm. (**G**)-(**L**) The corresponding Fourier transform of normalized dI/dV($r$, E) maps taken in (**A**)-(**F**). (**M**)-(**R**) The simulated QPI images at the corresponding energy considering two topological surface bands with band structures shown in Fig. 1**F**. (Detail in Note.S4).

**Fig. 4. Energy Dispersion of Topological Surface Bands by QPI.** The calculated constant energy contours (CECs) of two TSSs with spin texture at (**A**) E=950meV and (**B**) $E_F$. The black arrows represent the in-plane spin direction and the blue/red contours show the sign and the magnitude of out-of-plane spin. The color scale indicates the ratio of out-of-plane spin to total spin polarization. Blue and red arrows in each image indicate the relation between topological surface bands with scattering vectors, $q_1$ and $q_2$, respectively. The seemingly Rashba surface states are due to the unusual dispersion of the Dirac cone in Fig.1F, in which the outer and inner CEC (upper and lower part of Dirac cone) exist at the same energy. The outer TSS and the bulk state strongly overlap below E~300meV around $\overline{K}$ (gray regions in (B)), leaving small portion near $E_F$. The energy dispersion of (**C**) $q_1$ from E=500meV to E=1000meV and (D) $q_1$, $q_2$ and $q_3$ from E=-200meV to E=300meV. Hollow circles are extracted from the measured QPI images along $\overline{\Gamma M}$ direction, which are in excellent agreement with the QPI dispersion obtained from our calculated topological surface band structure (solid lines). We do not extract the exact $|q|$ in the energy range between 300meV and 500meV, in which blurry $q_1$ occurs due to flat band dispersion and larger <Sz> in TSS. $q_3$ is associated with the intraband scattering on hole-like bulk bands at $\overline{\Gamma}$.

**Fig. 5. Superconducting State and Vortex Imaging.** (**A**) The tunneling spectrum on Pb-terminated surface at T=0.26K (V=25mV, I=600pA and lock-in modulation 50μV). The black line is the result of a single BCS s-wave gap fitting (Detail in S.6). The inset shows the zoom-in of spectrum between 0.2meV and -0.2meV. (**B**) Temperature evolution of tunneling spectrum on Pb-terminated surface. Each spectrum is shifted by 10nS for clarity. (**C**) The temperature dependence of superconducting gap. Black line represents the theoretical temperature dependence from BCS theory. (**D**) Differential conductance map at E=0 shows the triangular vortex lattice, with applied magnetic field of 0.05T at T= 0.26K (V=10mV, I=60pA and lock-in modulation 250µV). (**E**) The tunneling spectrum at vortex core shows the structure of a vortex bound state (V=30mV, I=400pA, modulation=60µV, applied magnetic field of 0.05T and T=0.26K). (**F**) The spatial dependence of tunneling spectrum from the center of a vortex core toward the next nearest vortex core (V=30mV, I=600pA, lock-in modulation=60µV, the applied magnetic field of 0.05T and T=0.26K).

Figure 1

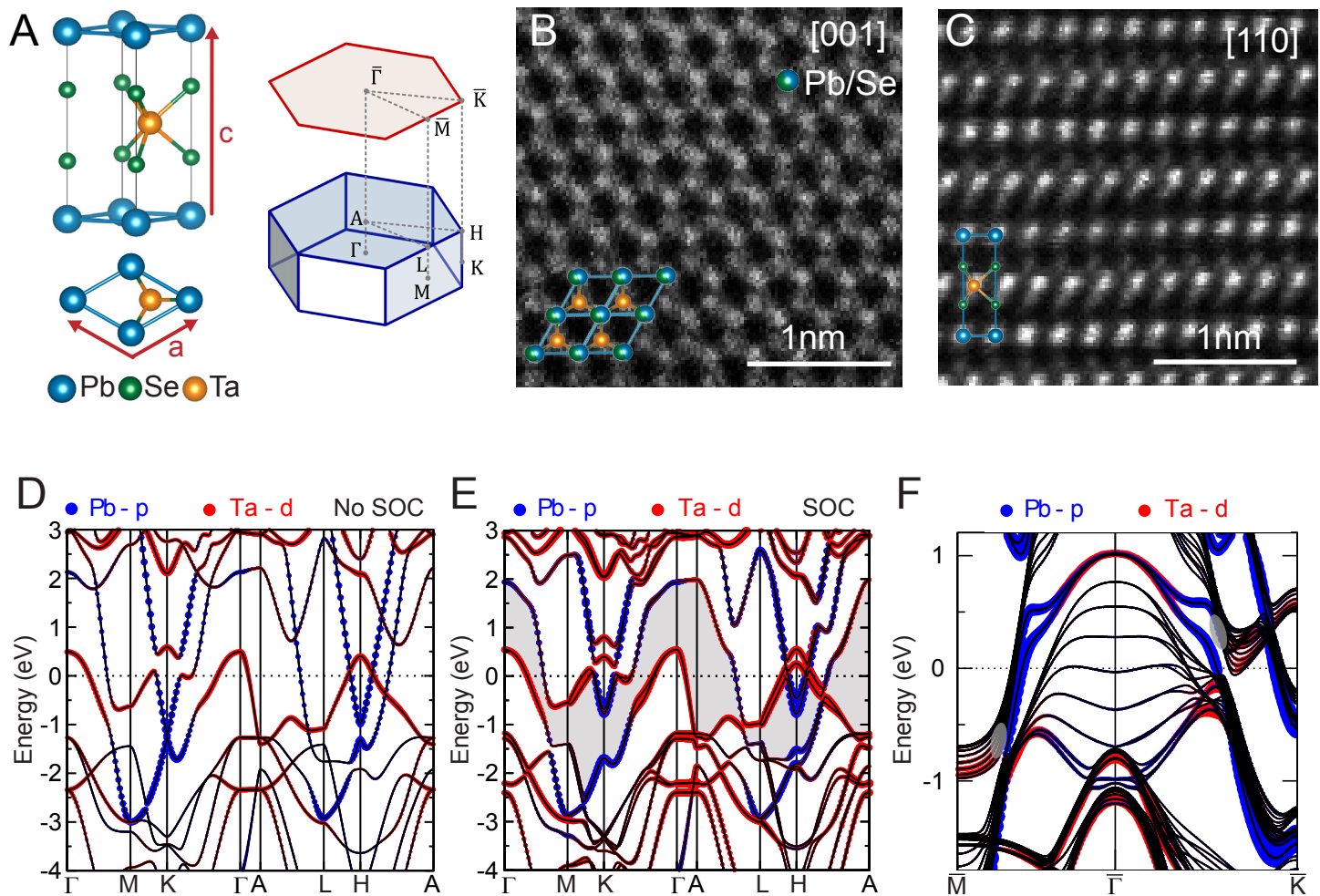

Figure 2.

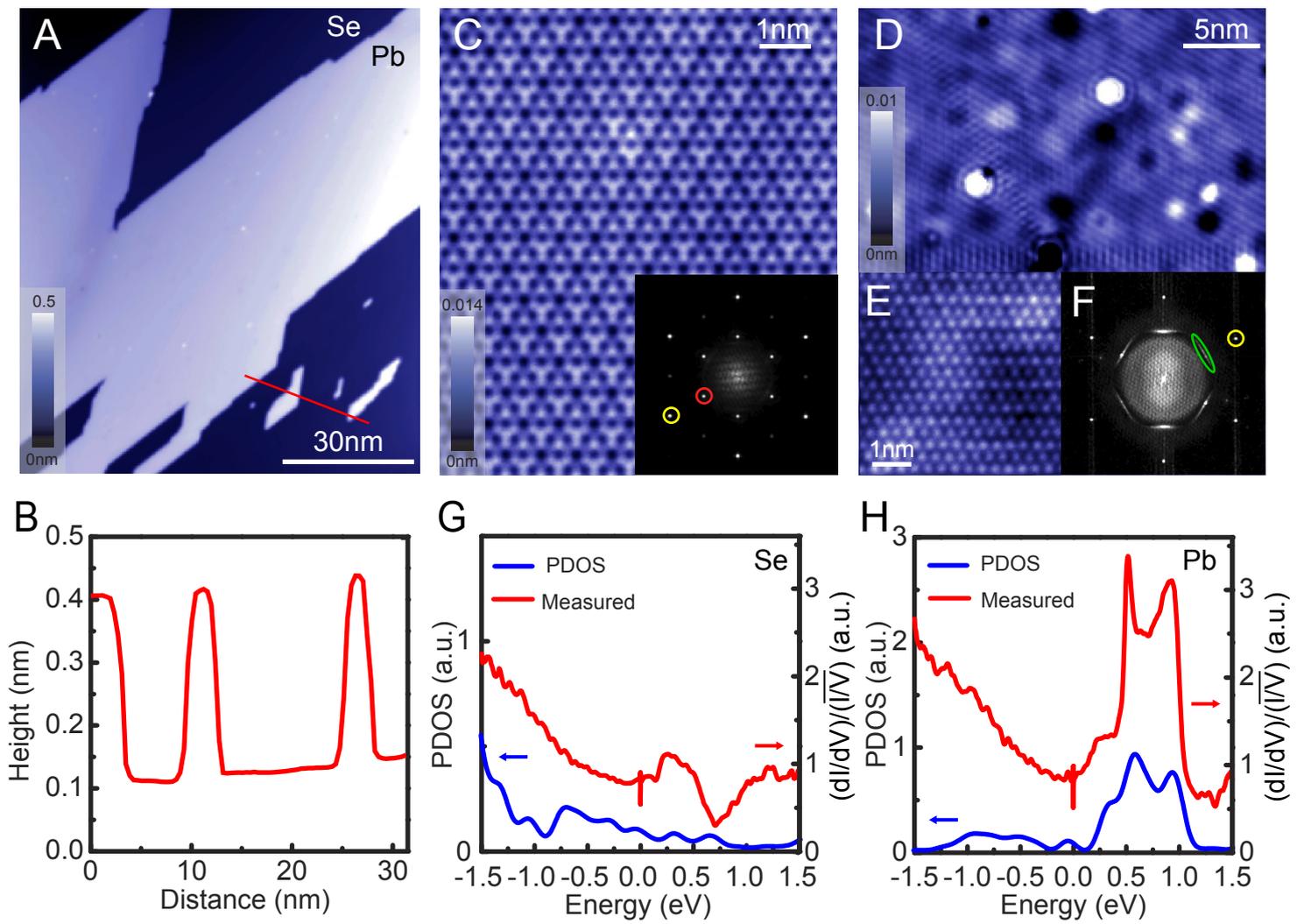

# Figure 3.

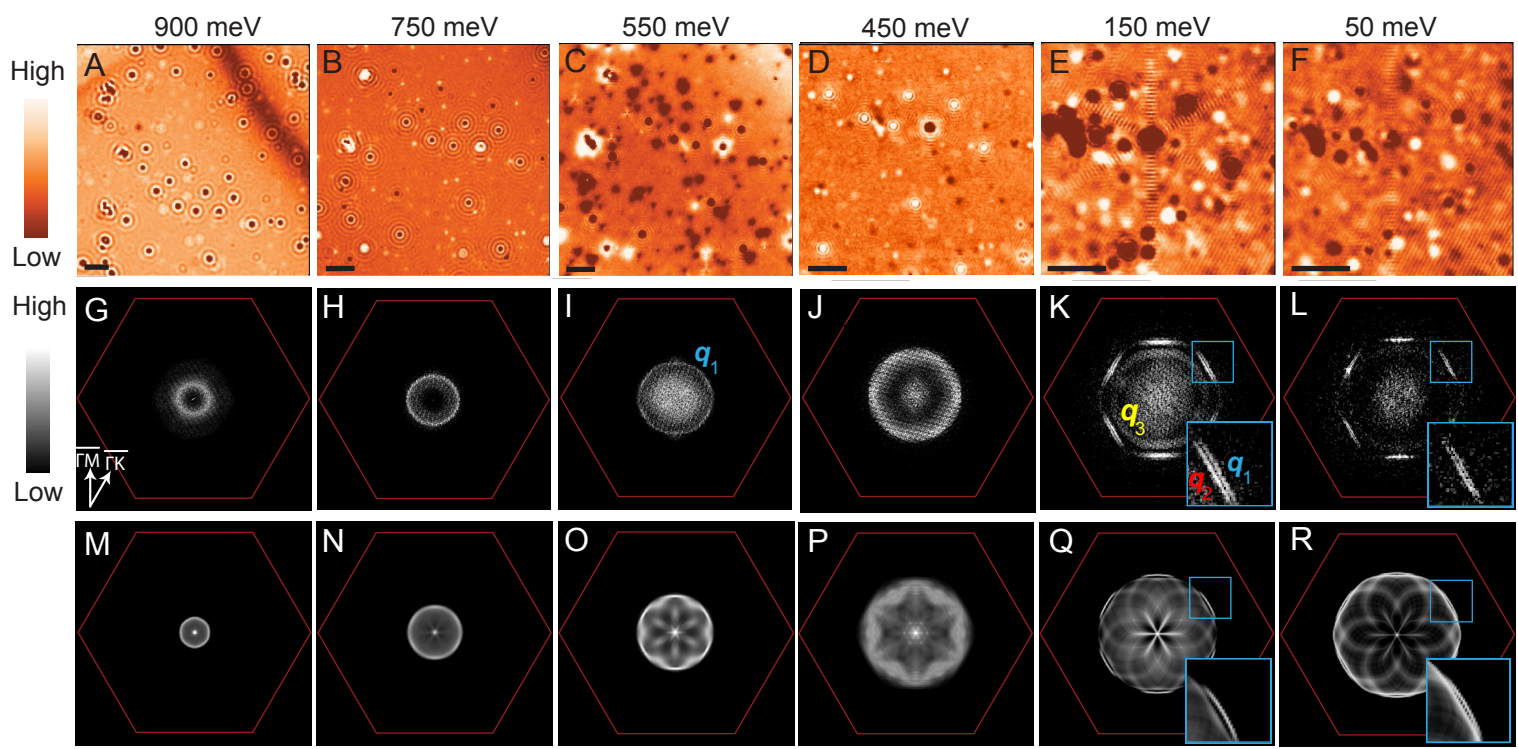

Figure 4.

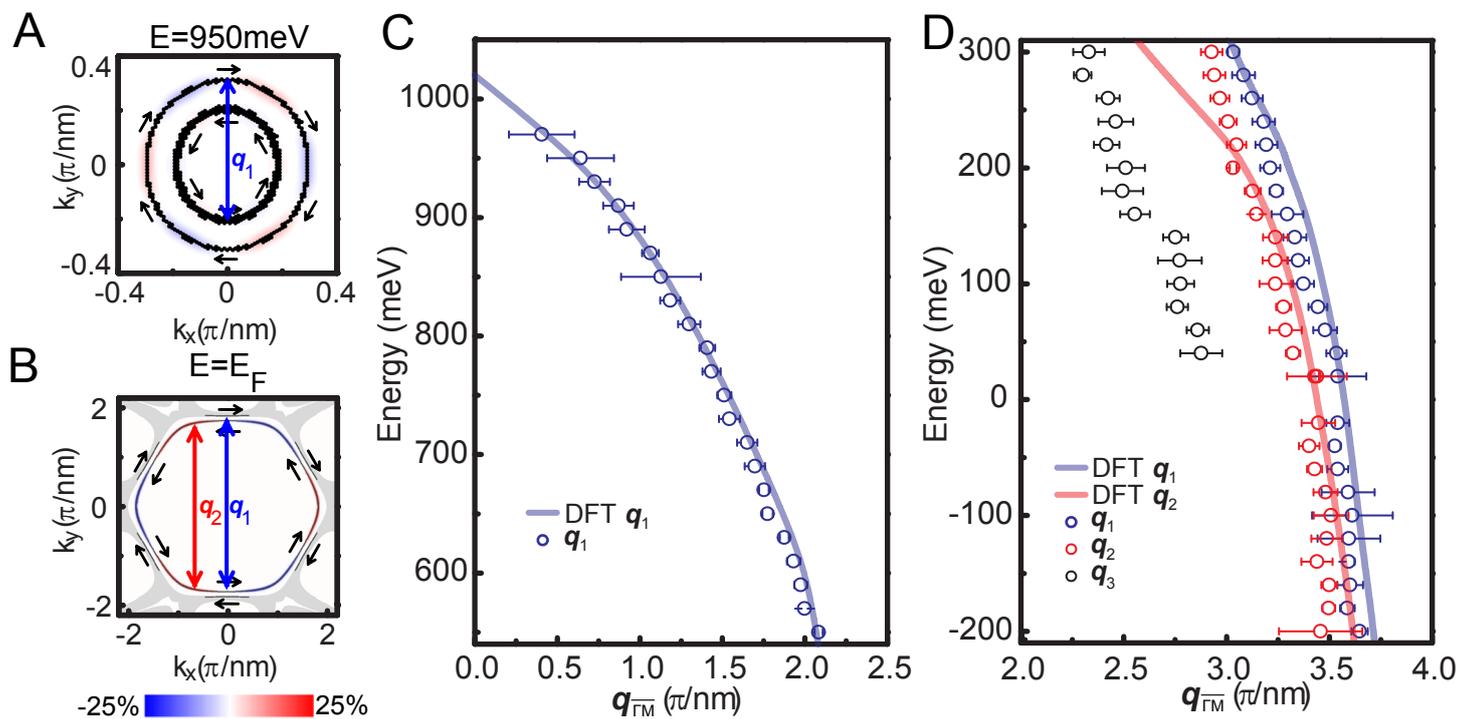

Figure 5.

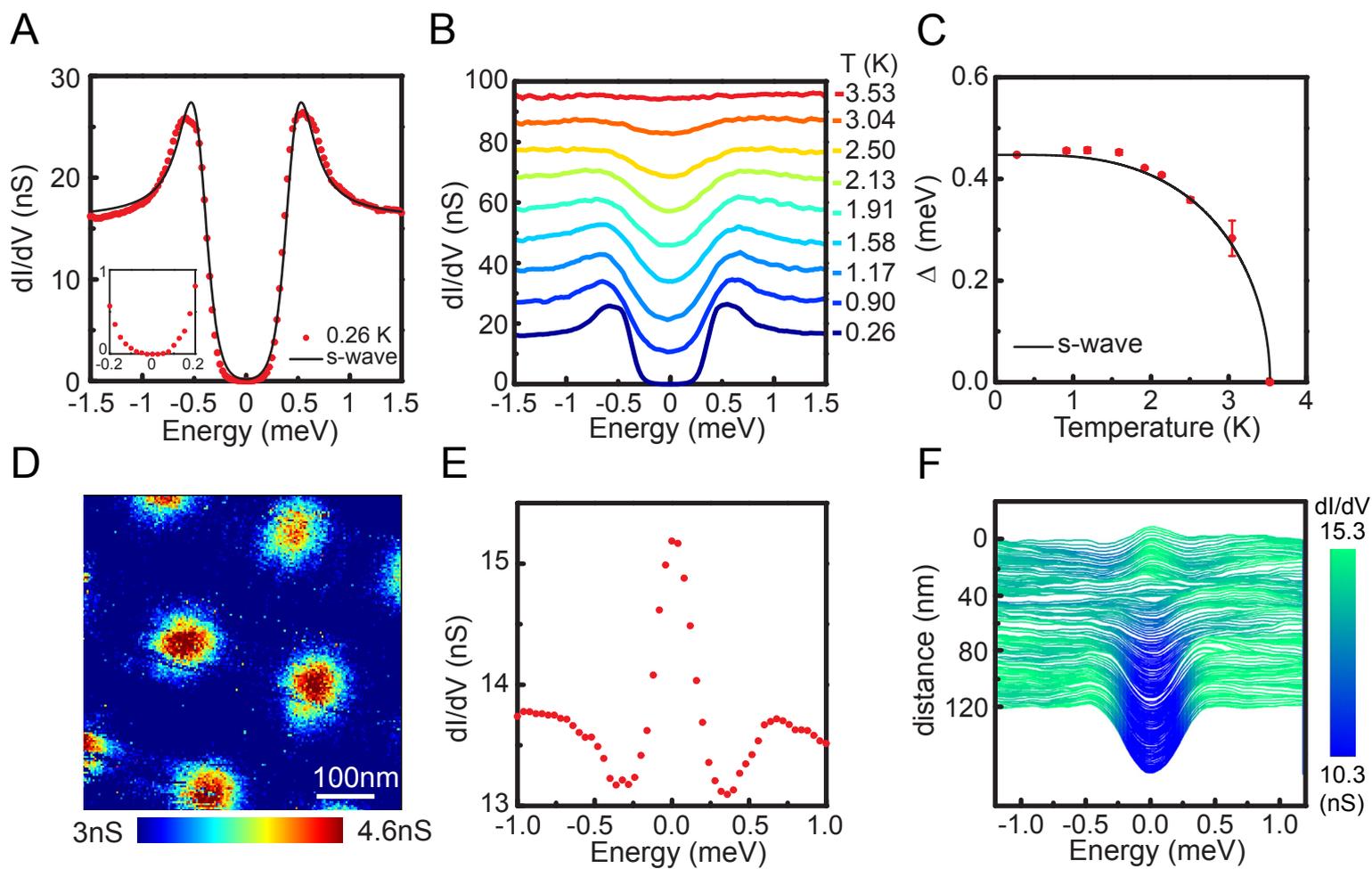